\documentclass[12pt,epsf]{article}
\usepackage{amsmath}
\usepackage{amsthm, amssymb}
\usepackage{graphicx}
\usepackage{setspace}
\usepackage{subfigure}
\usepackage{cite}
\usepackage{bm}
\usepackage{customcommands}

\def\href#1#2{#2}
\makeatletter
\def\namedlabel#1#2{\begingroup
    #2%
    \def\@currentlabel{#2}%
    \phantomsection\label{#1}\endgroup
}
\makeatother

\usepackage{color}

\usepackage{xcolor}
\usepackage[citebordercolor=green, linkbordercolor={ 0 0 1}, linktocpage=true]{hyperref}

\begin{document}
\begin{titlepage}
\begin{NoHyper}
\hfill
\vbox{
    \halign{#\hfil         \cr
           } 
      }  
\vspace*{20mm}
\begin{center}
{\Large \bf Causal Density Matrices}

\vspace*{15mm}
\vspace*{1mm}
Netta Engelhardt$^{a}$ and Sebastian Fischetti$^{b}$
\vspace*{1cm}
\let\thefootnote\relax\footnote{nengelhardt@princeton.edu, s.fischetti@imperial.ac.uk}\\
{$^{a}$Department of Physics, Princeton University\\
Princeton, NJ 08544, USA\\
\vspace{0.25cm}
$^{b}$Theoretical Physics Group, Blackett Laboratory, Imperial College \\ London SW7 2AZ, UK}

\vspace*{1cm}
\end{center}
\begin{abstract}
We define a new construct in quantum field theory -- the causal density matrix -- obtained from the singularity structure of correlators of local operators.  This object provides a necessary and sufficient condition for a quantum field theory state to have a holographic semiclassical dual causal geometry.  By exploiting the causal density matrix, we find that these dual causal geometries quite generally (even away from AdS/CFT) exhibit features of quantum error correction.  Within AdS/CFT, we argue that the ``reduced'' causal density matrix is the natural dual to the causal wedge. Our formalism is very well-suited to generalizations of holography beyond AdS/CFT or even gravity/QFT.
\end{abstract}
\end{NoHyper}

\end{titlepage}
\tableofcontents
\vskip 1cm
\begin{spacing}{1.2}

\setcounter{footnote}{0}


\section{Introduction}
\label{sec:intro}

The holographic principle~\cite{Tho93, Sus95, Bou02} posits that quantum theories of gravity are holographic: quantum gravity in~$(d+1)$ dimensions (hereafter referred to as the ``bulk'' theory) is dual to a lower-dimensional theory in~$d$ dimensions (the ``boundary'').  Within this framework, the \textit{states} of particular physical relevance are those in which the bulk can be described semiclassically\footnote{Theories of quantum gravity that do not admit \textit{any} such states are possibly interesting as theoretical artifacts, but hardly of relevance to a physical study.}, i.e.~the states that can describe the emergence of classical spacetime. What, then, are the lower-dimensional holographic duals of such states?

Because little is known about these duals in broad generality, we seek guidance from the most explicit manifestation of holography: the AdS/CFT correspondence~\cite{Mal97, GubKle98, Wit98a} (though ultimately our analysis in this paper will be quite general).  Even within this relatively well-understood example, the answer to our question remains elusive: while bulk states with a semiclassical description correspond to the large-$N$, large-$\lambda$ limit  of the dual local quantum field theory (QFT), the converse is manifestly false, as not all boundary states give rise to semiclassical gravity duals in this limit\footnote{Here, by ``semiclassical gravity'' we mean the usual definition of a system well-approximated by QFT on a curved background spacetime~$(M,g_{ab})$ (whose dynamics may be related to classical background matter fields, but backreaction from quantized fields can be ignored).} (though see e.g.~\cite{Van10} for early work on this question).  A key observation makes the problem more tractable:  the most fundamental necessary ingredient for a well-defined bulk QFT is not the bulk spacetime geometry~$(M,g_{ab})$ itself, but rather its \textit{causal structure}; it is precisely this structure that allows microcausality to be well-defined, a \textit{sine qua non} of local (relativistic) quantum field theory (bulk or boundary).  This observation motivates a simpler, more basic question: \\

\noindent \textit{Which states of a boundary theory holographically describe a bulk with semiclassical causal structure, and what precisely is this description?}\\

In this paper, we will answer this question most precisely when the boundary theory is a QFT, but we will also give a broad approach to answering the same question when the boundary theory is more general; we will then examine some of the consequences of the answer. Our most explicit construction applies to any holographic description of quantum gravity that obeys the following properties:
\begin{description}
	\item[\namedlabel{H1}{H1}]\textit{(Bulk)} When a semiclassical bulk exists, it contains some bulk matter field~$\phi(x)$ which is weakly interacting and thus can be treated perturbatively in the interaction coupling;
	\item[\namedlabel{H2}{H2}] \textit{(Boundary)} The lower dimensional theory is a QFT that lives on a timelike geometry that can be embedded on the boundary of the bulk (which need not be an asymptotic boundary);
	\item[\namedlabel{H3}{H3}] \textit{(Bulk-to-boundary)} There exists an ``extrapolate dictionary'' that relates correlators of local operators~$\Ocal(X)$ in the boundary to an appropriate limit of correlators of dual local operators~$\phi(x)$ in the bulk:
		\be
		\lim_{x_i \to X_i} \ev{\phi(x_1)\cdots\phi(x_n)}_\mathrm{bulk} = \ev{\Ocal(X_1)\cdots\Ocal(X_n)}_\mathrm{boundary},
		\ee
		where~$x_i$ and~$X_i$ denote points in the bulk and boundary, respectively.
\end{description}

We will call any duality obeying~\ref{H1}-\ref{H3} ``strong holography''; the large-$N$, large-$\lambda$ limit of AdS/CFT is one such example.  Since much of our discussion is guided by intuition from AdS/CFT, we find it useful for pedagogical purposes to sometimes frame it in that language. We emphasize, however, that our results are applicable to more general forms of holography; we will discuss later generalized versions of conditions~\ref{H1}-\ref{H3} (though we remain agnostic about the existence of such forms of holography). 

Before proceeding, it is worth pausing to address some potential objections to the question posed above.  First, as was pointed out in~\cite{Mar14}, emergent gravity is inherently nonlocal: it is described by a Hamiltonian that is purely a boundary term on shell.  This result may raise a potential concern regarding the validity of property~\ref{H1}: how can the bulk be local? This concern is unfounded, as our requirement is only \textit{approximate} bulk locality in the appropriate limit. This is precisely the case in e.g.~AdS/CFT, where semiclassically the bulk is well-described by local quantum fields on a classical asymptotically AdS spacetime. 

Second, advocates of the so-called strong form of AdS/CFT would contend that any state of a UV-complete CFT should have a bulk dual with asymptotically AdS boundary conditions.  Even if we adopt the idea that this strong form is true -- which is far from clear, especially when the boundary theory in question lives on a curved background, as allowed by our present discussion --  there is no tension with our question on the dual of semiclassical causal structure. We are interested in a \textit{sufficient} condition for the existence of a \textit{semiclassical} dual causal geometry and how this geometry is constructed, a question to which the strong form of AdS/CFT has no relevance. 

We now summarize the answer to our question -- ``when does a state of a QFT have a dual semiclassical causal structure?'' -- in the context of strong holography. We build on the work of~\cite{EngHor16a,EngHor16c} to construct a field theoretic object, which we term a \textit{causal state}~$\widetilde{\ket{\psi}}$ or \textit{causal density matrix}~$\tilde{\rho}$, that encodes precisely the boundary  information necessary to identify whether a dual bulk causal structure exists and (re)construct this dual if it does.  In other words, the causal state~$\widetilde{\ket{\psi}}$ is obtained from a physical state~$\ket{\psi}$ via a coarse-graining that discards any information not encoded in the dual causal structure. Our approach leads us to several attractive insights: for example, causal state holography appears to be quantum error correcting. It also answers a longstanding question in AdS/CFT: what is the dual of the so-called causal wedge~\cite{BouLei12, HubRan12}? Additional interesting features include defining ``entropy-like'' measures of coarse-graining, causal RG flow, and possible duals of bulk conformal invariants.

The relationship to the causal wedge deserves additional comment. Recall that due to superficial parallels between the causal wedge and the so-called entanglement wedge~\cite{CzeKar12, Wal12, HeaHub14}, some works have espoused the belief that the area of the rim of the causal wedge -- the so-called causal surface -- must be dual to an information-theoretic quantity in the boundary theory~\cite{HubRan12}.  However, the realization of those expectations via the conjecture of~\cite{FreMos13} was disproved in~\cite{KelWal13}, whose own conjecture, again motivated by intuitions about entanglement, was recently disproved in~\cite{EngWal17}.  More importantly, the general arguments of~\cite{EngWal17} make it clear that the geometric similarity between the causal wedge and the entanglement wedge is a red herring: the relationship between the area of the causal surface and the causal wedge is fundamentally different from the relationship between the area of the HRT surface~\cite{RyuTak06, HubRan07} and the entanglement wedge.  Attempts to draw na{\"i}ve insights about the causal wedge by drawing on intuitions regarding the entanglement wedge are fundamentally misguided; the two should not be thought of as analogous objects.

It is clear that a paradigm shift in the general approach to the causal wedge dual is required. Here we advocate a way of thinking about the causal wedge that does not rely upon entanglement-based intuition. We do not assume that the area of the causal surface is a special quantity, and we do not endorse the idea that it should have a simpler interpretation than the area of any other surface.  Instead, we adopt the following stance: since the causal wedge is defined purely from the \textit{causal structure} of the bulk, only the causal structure of the causal wedge should have a natural holographic dual. This dual, as we will show, is none other than the (reduced) causal state~$\widetilde{\ket{\psi}}_\Rcal$.

A potential objection to this viewpoint, of course, comes from the argument that the area of the causal surface should be ``distinguished'' if the Generalized Second Law~\cite{Bek73, Haw76} is to have a meaningful holographic interpretation on the boundary.  Moreover, since the causal surface lies in the entanglement wedge, its area is obviously computable from the reduced density matrix~\cite{JafLew15, DonHar16}.  These considerations do not, however, motivate that the area, or more generally, the generalized entropy, of the causal surface should be special \textit{from the perspective of the causal wedge}.  Our opinion is that the area of the causal surface does not have a nice information-theoretic interpretation in terms of only data necessary to reconstruct the causal wedge; we do not exclude the possibility of a nice interpretation in terms of the full data necessary to reconstruct the entanglement wedge.

\subsection*{Summary of Results}
\begin{figure}[t]
\centering
\subfigure[]{\includegraphics[width=4cm,page=1]{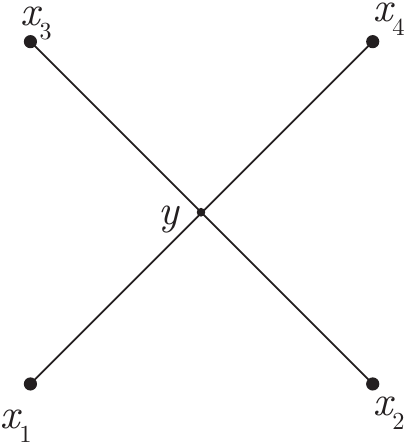}
\label{subfig:generalLandau}
}
\hspace{2cm}
\subfigure[]{\includegraphics[width=3cm,page=2]{Figures-pics.pdf}
\label{subfig:bndryLandau}
}
\caption{\subref{subfig:generalLandau}: In a general QFT with perturbative interactions, the correlator~$\ev{\phi(x_1)\cdots\phi(x_n)}_\psi$ is singular when all the~$x_i$ are null separated from a common vertex~$Y$, with the corresponding Landau diagram (shown here) obeying all conservation laws.  This is the so-called ``lightcone singularity''. \subref{subfig:bndryLandau}: Reproduced from~\cite{EngHor16a}. From the point of view of the boundary field theory, these bulk-point singularities of a \textit{boundary} correlator~$\ev{\Ocal(X_1) \ldots \Ocal(X_n)}_\psi$ identify a bulk point~$p$.} 
\label{fig:Landau}
\end{figure}

Let us give a brief preview of how Properties~\ref{H1}-\ref{H3} define a causal state and the corresponding dual causal structure.  Consider a state~$\ket{\psi}$ with a semiclassical dual causal structure in the sense described above, and consider the~$n$-point function $\ev{\phi(x_1)\cdots\phi(x_n)}_\psi$ of some local quantum bulk field~$\phi(x)$.  If~$\ket{\psi}$ is Hadamard (which we will assume throughout), then this correlator exhibits two types of singularities: when two or more of the points~$x_i$ are coincident, or when all of the~$x_i$ are null separated from a common point~$y$ with the corresponding Landau diagram obeying all conservation laws, as shown in Figure~\ref{subfig:generalLandau}.  

The extrapolate dictionary dictates that, in the limit that the points $x_{i}$ are taken to boundary points $X_{i}$, the boundary~$n$-point correlator~$\ev{\Ocal(X_1) \ldots \Ocal(X_n)}_\psi$ inherits the singularity structure of~$\ev{\phi(x_1)\cdots\phi(x_n)}_\psi$. From a purely \textit{boundary} perspective, the correlator~$\ev{\Ocal(X_1) \ldots \Ocal(X_n)}_\psi$ now exhibits three types of singularities: the usual universal short-distance singularities, the usual universal light-cone singularities due to null separation on the boundary, and non-universal singularities when all the~$X_i$ are null-separated from a \textit{bulk} point~$p$ when a semiclassical bulk exists, as shown in Figure~\ref{subfig:bndryLandau}.  These latter singularities were coined ``bulk-point singularities'' in~\cite{MalSimZhi}, and their existence, investigated earlier in~\cite{PolSus99, GarGid09, HeePen09, Pen10, OkuPen10}, is a necessary condition for the existence of a semiclassical holographic dual causal structure (with perturbative dynamics).  This condition alone, however, is not sufficient to ensure the existence of a dual semiclassical conformal geometry. How are we to determine whether these ``extra'' singularities in~$\ev{\Ocal(X_1) \ldots \Ocal(X_n)}_\psi$ are sourced by a well-behaved semiclassical bulk or are just artifacts of some strange, possibly pathological aspect of the field theory?

The answer was provided in~\cite{EngHor16a,EngHor16c}, where it was shown that -- when a semiclassical bulk exists -- the singularities in~$(d+3)$-point correlators (with~$d$ the dimension of the boundary spacetime) can be used to define special spacelike slices of the boundary spacetime; these were called \textit{lightcone cuts} because they correspond to the intersection of lightcone of~$p$ with the boundary~\cite{New76}.  The geometry of these cuts yields an overdetermined system of algebraic equations for the conformal metric\footnote{That is, the metric up to an overall conformal factor: this object is precisely the causal structure.} at~$p$ of the bulk dual; by construction, these equations have a unique consistent solution whenever the singularities in question are sourced by a local, causal bulk.

This boundary-to-bulk procedure may be made precise as follows: consider the~$(d+3)$-point correlator~$\ev{\Ocal(X_1) \ldots \Ocal(X_{d+3})}_\psi$ of some local operator~$\Ocal(X)$ in the boundary theory.  If the construction of~\cite{EngHor16a,EngHor16c} from the singularity structure of this correlator gives rise to a well-defined conformal geometry, then that geometry \textit{is} the semiclassical holographic dual causal structure.  Moreover, there exists a perturbatively interacting quantum field~$\phi(x)$ dual to~$\Ocal(X)$ which gives rise to the singularities in~$\ev{\Ocal(X_1) \ldots \Ocal(X_{d+3})}_\psi$. This is explicit spacetime emergence.

The causal state~$\widetilde{\ket{\psi}}$ advertised above is designed to capture precisely this singularity structure by coarse-graining over everything else.  Roughly, we say that two states~$\ket{\psi}$ and~$\ket{\psi'}$ are causally equivalent, denoted~$\ket{\psi}\sim\ket{\psi'}$, if they produce the same singularity structure in the correlator~$\ev{\Ocal(X_1) \ldots \Ocal(X_{d+3})}$.  Then the causal state~$\widetilde{\ket{\psi}}$ associated to some state~$\ket{\psi}$ is precisely the equivalence class of~$\ket{\psi}$ under~$\sim$.  In fact, we will argue that it is possible to quotient the full Hilbert space of the theory by~$\sim$, thereby producing a \textit{causal Hilbert space}~$\widetilde{\Hcal}$. The causal state is an element of~$\widetilde{\Hcal}$, which allows definition of the causal density matrix~$\tilde{\rho} \equiv \widetilde{\ket{\psi}} \widetilde{\bra{\psi}}$ as a linear operator on~$\widetilde{\Hcal}$.  By construction, therefore, all that is needed to determine the semiclassical holographic dual causal structure of a state~$\ket{\psi}$ (and whether or not there even is one) is the causal state~$\widetilde{\ket{\psi}}$.

We find it useful to further consider a generalization of the causal state $\widetilde{\ket{\psi}}$ (or density matrix~$\tilde{\rho}$): given a subregion~$\Rcal$ of the boundary spacetime, we may restrict our attention to the portion of lightcone cuts that intersects~$\Rcal$.  By performing this restriction to~$\Rcal$ in two distinct ways (one of which is a stricter version of the other), we obtain two insights.

First, if we require that the correlator~$\ev{\Ocal(X_1) \ldots \Ocal(X_{d+3})}$ be singular when \textit{all} of the~$X_i$ are contained in~$\Rcal$, we find two intriguing features: \textit{(i)} there is redundancy in how the conformal metric at a particular bulk point~$p$ is encoded in the singularity structure of~$\ev{\Ocal(X_1) \ldots \Ocal(X_{d+3})}$; \textit{(ii)} when~$\Rcal$ is too small, the singularity structure of~$\ev{\Ocal(X_1) \ldots \Ocal(X_{d+3})}$ restricted to~$\Rcal$ is insufficient to reconstruct the conformal metric at~$p$.  These two features bear a strong resemblance to recent insights on \textit{(i)} quantum error correction and \textit{(ii)} quantum secret-sharing in AdS/CFT~\cite{AlmDon15, HaPPY, Har16, MinPol15}; our result may be viewed as a form of ``causal'' quantum error correction in a more general holographic setting than just AdS/CFT.

\begin{figure}[t]
\centering
\includegraphics[width=2.5cm,page=3]{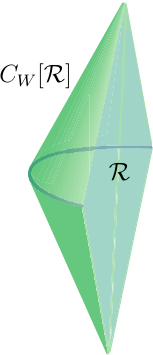}
\caption{The causal wedge $C_{W}[\Rcal]$ of some causal boundary region~$\Rcal$ is  the intersection of the past and future of~$\Rcal$ in the bulk.}
\label{fig:causalwedge}
\end{figure}

Second, consider the case where the correlator~$\ev{\Ocal(X_1) \ldots \Ocal(X_{d+3})}$ is singular when only \textit{some} the~$X_i$ are required to be contained in~$\Rcal$.  These singularities are sufficient to reconstruct the portion of the lightcone cuts that intersect~$\Rcal$, which in turn identify bulk points that lie in the intersection of the past and future of~$\Rcal$.  When~$\Rcal$ is globally hyperbolic (colloquially known as a boundary causal diamond), this intersection is the causal wedge~$C_{W}[\Rcal]$ of~$\Rcal$, as shown in Figure~\ref{fig:causalwedge}.  By an appropriate additional coarse-graining, we may define a \textit{reduced} causal state on~$\Rcal$,~$\widetilde{\ket{\psi}}_\Rcal$, which encodes the information necessary to reconstruct the conformal geometry of~$C_{W}[\Rcal]$.  The amount of coarse-graining performed in going from~$\ket{\psi}$ to~$\widetilde{\ket{\psi}}$ can be captured by a number~$D_\psi$ which roughly encodes the ``size'' of the subspace of~$\Hcal$ that projects to~$\widetilde{\ket{\psi}}$; this object is similar in spirit to (but very different in construction from) the entanglement entropy.

We now discuss more general forms of holography. Motivated in part by the subtleties involved in the formulation of a holographic duality of two (approximately) local QFTs, we also consider more general dualities that obey a partial relaxation of ``strong holography''.  We find that much of our discussion generalizes, albeit less explicitly, to any system obeying Property~\ref{H1} above and weaker versions of Properties~\ref{H2} and~\ref{H3}:

\begin{description}
	\item[\namedlabel{H2prime}{H2$'$}] \textit{(Boundary)} The lower-dimensional theory is a well-defined dynamical theory that can be embedded on the boundary of the bulk (which need not be an asymptotic boundary). This boundary is timelike or null;
	\item[\namedlabel{H3prime}{H3$'$}] \textit{(Bulk-to-boundary)} There exists an ``extrapolate dictionary'' that relates correlators of local bulk fields $\phi(x)$ to \textit{some} object $O(X)$ in the boundary theory in the appropriate limit:
		\be
		\label{eq:Odef}
		\lim_{x_i \to X_i} \ev{\phi(x_1)\cdots\phi(x_n)}_\mathrm{bulk} = O(X_{1}\cdots X_{n})_\mathrm{boundary},
		\ee
		where~$x_i$ and~$X_i$ denote points in the bulk and boundary spacetimes, respectively.
\end{description}
Because the left-hand side of equation~\eqref{eq:Odef} encodes bulk scattering, the object $O(X)$ is ``morally'' the bulk S-matrix; any reasonable holographic dual should therefore obey~\ref{H3prime}.

We call a duality obeying properties~\ref{H1},~\ref{H2prime}, and~\ref{H3prime} ``weak holography''  (see also~\cite{NomSal16b} for work on a formulation of general holography).  Many of our statements apply equally well to such systems, provided the objects $O(X)$ in the boundary theory are known.  For most of this paper, we will specialize to the strong form of holography, where~$O(X)$ is the expectation value of a composite operator; in Section~\ref{sec:hol} we will examine which of our statements are generalizable to weak holography.

This paper is structured as follows.  In Section~\ref{sec:CDM}, we review and generalize the construction of~\cite{EngHor16a,EngHor16c} before defining precisely the equivalence relation~$\sim$ and the causal state and causal density matrix.  In Section~\ref{sec:QEC} we discuss the similarities between our construction and quantum error correction. In Section~\ref{sec:causalwedge} we discuss the natural interpretation of the causal wedge that emerges from our formalism.  Section~\ref{sec:hol} gives a generalization of our results for weak holography. We conclude in Section~\ref{sec:disc} with some discussion and future directions.

\paragraph{Assumptions and Conventions:} We assume that the boundary theory lives on a~$C^{2}$ maximally extended,~$d$-dimensional, globally hyperbolic manifold $\partial M$ that is either timelike or null and geodesically complete (or, if this is a conformal boundary, it can be put in a conformal frame that is geodesically complete).  We will restrict our attention to emergent conformal geometries on manifolds that are $C^{2}$, maximally extended, connected,~$(d+1)$-dimensional, and globally or AdS hyperbolic~\cite{Wal12}.  The condition of global or AdS hyperbolicity in the bulk is convenient but not essential; we will comment on it in Section~\ref{sec:disc}.  All correlation functions are assumed to be time-ordered. All other conventions are as in~\cite{Wald} unless otherwise stated.

\section{The Causal Density Matrix}
\label{sec:CDM}

In this section, we describe precisely how the causal state~$\widetilde{\ket{\psi}}$ is defined, beginning with a review of the lightcone cut construction.  We will restrict to the case of strong holography; the generalization to weak holography will be presented in Section~\ref{sec:hol}.

\subsection{The Lightcone Cut Construction}
\label{subsec:cutreview}

In~\cite{EngHor16a, EngHor16c}, it was shown (in the context of AdS$_{d+1}$/CFT$_d$) that the singularity structure of~$(d+3)$-point CFT correlators can be used to obtain the causal structure of a semiclassical bulk gravitational dual.  Note that while the construction of~\cite{EngHor16a, EngHor16c} restricted itself to the context of AdS/CFT, its ingredients apply more generally.  The purpose of this subsection is both to review and generalize the salient results of~\cite{EngHor16a}. We will begin with the geometrical (bulk) aspects of the construction, and then relate them to the singularities of~$(d+3)$-point correlators in the dual field theory.

To that end, recall that the future (past) of a point~$p$, denoted~$I^{+}(p)$~($I^{-}(p)$), is defined as the set of all points that can be reached from~$p$ via a future-directed (past-directed) timelike path. The boundary of the future (past), denoted~$\partial I^{+}(p)$~($\partial I^{-}(p)$), is an achronal surface generated by null future-directed (past-directed) geodesics from~$p$.  These generators leave the surface after caustics and intersections, ensuring achronality. 

The future \textit{lightcone cut} of a point $p$, denoted $C^{+}(p)$, is the intersection of the boundary of the future of $p$ with the boundary, with a similar definition for the past lightcone cut $C^-(p)$:
\be 
C^{\pm}(p)= \partial I^{\pm}(p)\cap \partial M.
\ee
We will use $ C(p)$ as shorthand whenever it does not matter if we are discussing the past or future lightcone cut of $p$, and we will call it ``the cut of $p$'' for short.

\begin{figure}[t]
\centering
\includegraphics[width=4.5cm,page=4]{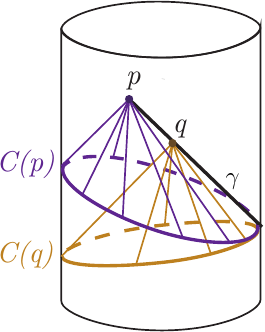}
\caption{The lightcone cuts of $p$ and $q$ are tangent due to a shared generator $\gamma$ (thick black line) of $\partial J(p)$ and $\partial J(q)$.}
\label{fig:nullsep}
\end{figure}

We will need to extend several key results from~\cite{EngHor16a}. First, it was proven in~\cite{EngHor16a} that points $p$ in the causal wedge of $\partial M$,  defined as~$C_{W}[\partial M] = \partial I^{+}[\partial M]\cap \partial I^{-}[\partial M]$, are in a one-to-one correspondence with lightcone cuts $C^{\pm}(p)$. Thus the causal wedge of $\partial M$ can be represented as the space of past and future lightcone cuts.  Second, it was shown that if two lightcone cuts $C(p)$ and $C(q)$ are tangent at a (boundary) point $X$, then the corresponding bulk points $p$ and $q$ are null-separated\footnote{By null-separated, we mean that there exists a null curve connecting them, but no timelike one.}; see Figure~\ref{fig:nullsep} for an illustration. In~\cite{EngHor16a}, it was assumed that $ M$ was asymptotically AdS (and thus $\partial M$ was conformal to the Einstein Static Universe). We extend these results below in four propositions to include any boundary spacetime obeying our assumptions.

\paragraph{Proposition 1: \label{prop1}} If $C(p)$ and $C(q)$ are tangent at a point $X$, then $p$ and $q$ are null-separated. 
\begin{proof} Let us work with past cuts for simplicity.
Since $C^{-}(p)$ is by assumption $C^{1}$ at $x$, then $\partial I^{-}(p)$ must $C^{1}$ at $X$ as well. There is therefore a unique generator of $\partial I^{-}(p)$ at $X$, and this unique generator is orthogonal to $C^{-}(p)$ at $X$. Because $\partial M$ is causal, there is precisely one future-directed inwards-going null vector orthogonal to~$C^-(p)$ at~$X$. This is the generator $\gamma$ of $\partial I^{-}(p)$ at $X$. But the statements above hold equally well for $\partial I^{-}(q)$, so $\gamma$ is also a generator of $q$. Therefore $\gamma$ goes through both $p$ and $q$. Because $\gamma$ is a generator of both $\partial I^{-}(p)$ and $\partial I^{-}(q)$ and is achronal, $p$ and $q$ must be null-separated. 
\end{proof}

\paragraph{Proposition 2: \label{prop2}} $C(p)$ is acausal whenever $p\in M$. 
\begin{proof}
By global (or AdS) hyperbolicity, $C(p)$ is achronal. If $\partial M$ is timelike everywhere on its intersection with $C(p)$, the result follows trivially as well: the intersection of a timelike surface with an achronal surface is spacelike. Suppose $\partial M$ is null somewhere on its intersection with $C(p)$. Then $C(p)$ must be spacelike unless $\partial I(p)$ shares a generator with $\partial M$ (this generator does not leave the congruence due to intersections). But then $p\in\partial M$. 
\end{proof}
 
\paragraph{Proposition 3:\label{prop3}} $C(p)$ is a complete continuous spatial slice of any connected component of $\partial M$ that intersects $I(p)$.  Moreover,~$C(p)$ is~$C^1$ on all but a measure-zero set. \\

\noindent The proof of this is identical to the proof in~\cite{EngHor16a}, thanks to Proposition 2.

Propositions 1-3 immediately allow us to determine the sign of null-separation: if $C(p)$ and $C(q)$ are tangent at a point $X$, and in a neighborhood of $X$ $C(p)$ is in the causal past of $C(q)$, then there is a future-directed null geodesic from $p$ to $q$. 

\paragraph{Proposition 4:\label{prop4}} Let $p\in I^{+}[\partial M]$. Then~$(i)$ $C^{-}(p)$ is nonempty, and~$(ii)$ if $C^{-}(p)=C^{-}(q)$, then $p=q$. A similar result holds for future cuts. 

\begin{proof}
$p\in I^{+}[\partial M]$ implies that $I^{-}(p)\cap \partial M \neq \varnothing$. Global (AdS) hyperbolicity implies that $\exists \ X \in \partial M$ such that $X\notin  I^{-}(p)$. Therefore $\partial I^{-}(p)\cap \partial M \neq \varnothing$, proving~$(i)$.  Next, suppose $C^{-}(q)=C^{-}(p)$. Then by Rademacher's theorem~\cite{Rademacher1919} (and the fact that the lightcone cuts are Lipshitz manifolds~\cite{HawEll}), $C^{-}(p)$ and $C^{-}(q)$ agree on a $C^{1}$ open set. Therefore $\partial I^{-}(p)$ and $\partial I^{-}(q)$, by the reasoning of the proof of Proposition 1, share an open set of generators. By global (AdS) hyperbolicity, $p=q$, proving~$(ii)$. 
\end{proof}

Note that because the $\partial I(p)\cap \partial M$ is a spacelike slice of the geometry (for causal boundaries) by Property 3, this also means that two different lightcone cuts cannot correspond to one point (they would be causally separated from one another).

The final result that we will use is a prescription for obtaining the conformal metric at a point $p$, i.e. the metric up to an overall function, from a sufficiently large discrete set of null vectors at $p$. This remains unaltered from~\cite{EngHor16a}. The idea is simple: if $\{\ell_{i}\}_{i=1}^{d+1}$ is a known set of $(d+1)$ linearly independent null vectors, then the metric at $p$ is determined, up to an overall factor\footnote{If the $\ell_{i}$ were not null, the metric would be fully determined.}, by the inner products of the $\ell_{i}$. By assumption, the $\ell_{i}$ are null, so $\ell_{i}\cdot \ell_{i}=0$ (no summation on the $i$ index). To fix the conformal metric, we need to determine the inner products $\ell_{i}\cdot \ell_{j}$ for $i\neq j$. Let $\{\eta_{k}\}$ be another known set of null vectors at $p$, which can be expanded in the $\ell_{i}$ basis:
\be
\eta_{k}=\sum\limits_{i}M_{ik}\ell_{i}.
\ee
Because the $\eta_{k}$ are null, we know that the corresponding inner products vanish. This yields a set of algebraic equations for the inner products $\ell_{i}\cdot \ell_{j}$:
\be \label{eq:sys}
0=\eta_{k}\cdot \eta_{k}=\sum\limits_{i,j}M_{ik}M_{jk}\ell_{i}\cdot \ell_{j}.
\ee
Equation~\eqref{eq:sys} determines the metric at $p$ up to an overall factor in terms of the known coefficients~$M_{ij}$. These equations are generically overdetermined, but we are guaranteed a solution whenever there exists a well-defined bulk metric at $p$. 

However, we are of course interested in determining the conformal metric at $p$ from \textit{boundary} data rather than bulk data. Using the approach above then requires being able to express null vectors at a bulk point $p$ in terms of objects defined on the boundary. This is precisely what the lightcone cuts do: if $C(p)$ and $C(q)$ are tangent at a point, then there is a null geodesic connecting them, which corresponds to a null vector at $p$. 

More explicitly, using the cut-point correspondence, we may work in the space of lightcone cuts ${\cal M}$ instead of the bulk: a lightcone cut $C(p)$ is a point $P$ in the space of cuts ${\cal M}$. We \textit{define} $P$ and $Q$ to be null-separated if $C(p)$ and $C(q)$ are tangent, which is the same as defining $P$ and $Q$ to be null-separated when $p$ and $q$ are null-separated. Thus we have endowed ${\cal M}$ with the same Lorentzian structure as the bulk $M$. By taking enough lightcone cuts $C(q)$ that are tangent to $C(p)$, we generate enough null vectors in the lightcone of $P$ to write down the system of equations~\eqref{eq:sys} for the conformal metric at $P$. Because ${\cal M}$ has the same conformal metric as $M$, this is equivalently a system of equations for the conformal metric at $p$. Whenever the lightcone cuts are generated by a well-defined bulk metric, there will be a unique, consistent solution. 

\begin{figure}[t]
\centering
\includegraphics[width=5cm,page=5]{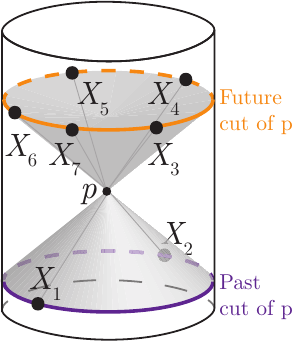}
\caption{The lightcone cuts of a point $p$ constructed from its bulk-point singularities.}
\label{fig:vertex}
\end{figure}

In order to complete the reconstruction of the conformal metric from boundary data, we must determine how to obtain the lightcone cuts from boundary data.  This can be accomplished by noting that the lightcone cuts $C^{\pm}(p)$ of a bulk point $p$ are precisely the boundary locations that are null (and achronally) separated from a single bulk point $p$. Thus, if $\{X_{i}\}_{i=1}^{n}$ is a set of $n$ points on $C^{\pm}(p)$, then they correspond to the minimal time separation at which we can draw a bulk Landau (position-space) diagram for some perturbatively-interacting bulk quantum field $\phi$ such that the diagram endpoints are all null-separated from the common bulk vertex $p$ (see Figure~\ref{fig:vertex}). When this Landau diagram is on-shell, it produces a ``light-cone singularity'' in the $n$-point correlator of the perturbatively interacting field $\phi$.

The extrapolate dictionary immediately implies that the time-ordered $n$-point correlator of the local boundary operator~$\Ocal(X)$ dual to $\phi(x)$ is also singular~\cite{MalSimZhi}. Thus the lightcone cuts are constituted of singularities of these boundary correlators.

We may use this observation to construct the full cuts from correlators as follows.  First, note that a set of $(d+1)$ null achronal geodesics in a $(d+1)$-dimensional bulk uniquely identify a bulk point if they intersect; thus to identify a bulk point, we only need a singular $(d+1)$-point correlator. However, generically $(d+1)$-point correlators corresponding to a diagram with a bulk point vertex are not singular, as the bulk Landau diagram does not conserve energy-momentum at the vertex. To ensure conservation of energy-momentum,~\cite{MalSimZhi} considered instead $(d+2)$-point correlators. Thus, to uniquely fix a bulk point, we look for $(d+2)$-points that result in singular correlators. To find the lightcone cuts,~\cite{EngHor16a} added one more point: a $(d+3)$-point correlator gives the freedom of moving one point to trace out the lightcone cut while moving another point to conserve energy momentum. Keeping $(d+1)$ points fixed guarantees that the corresponding bulk point does not move around as the lightcone cut is traced out. An additional requirement of~\cite{EngHor16a} was that the singularity of the $(d+3)$-point correlator be at the smallest possible time-separation on the boundary. This ensures that the object being traced out is the actual lightcone cut, rather than some other cross-section of the lightcone; this subtlety is due to geodesic intersections.  

The above construction is restricted to the boundary's causal wedge, as it requires conservation of energy-momentum at bulk points, though it needs a small modification in cases of collapsing black holes.  In generic spacetimes, it permits the determination of the cuts of points anywhere in the causal wedge, but in certain nongeneric situations like eternal black holes, conservation of energy-momentum at the vertex excludes components of the causal wedge as well.

To summarize the construction: consider any local operator~$\Ocal(X)$; if its time-ordered~$(d+3)$-point correlator has a singularity structure that consistently defines a causal geometry by equation~\eqref{eq:sys}, then that causal geometry is the emergent dual, and the operator~$\Ocal(X)$ is dual to some perturbatively interacting, local, causal dynamical field on it.

%
%


\subsection{Causal States}
\label{subsec:CDM}

The lightcone cut construction reviewed above gives an explicit algorithm for generating an emergent causal structure from QFT data in the form of the singularities of time-ordered~$(d+3)$-point functions.  \textit{A priori}, this data alone does not recover the entire bulk geometry, as it does not capture any information about the bulk conformal factor\footnote{It also by definition misses any spacetime inside an event horizon, and as discussed in~\cite{EngHor16a}, potentially also regions outside of event horizons in nongeneric spacetimes.}; thus we can think of the data contained in this singularity structure as a type of coarse-graining over the bulk conformal factor.  This is the coarse-graining promised in Section~\ref{sec:intro}, which we will now use to define causal states.

To develop some intuition, consider a (pure) state~$\ket{\psi}$ of the field theory.  The key observation is that the conformal geometry dual to~$\ket{\psi}$ is encoded in just the singularity structure of~$O_\psi(X_i) \equiv \ev{\Ocal(X_1) \cdots \Ocal(X_{d+3})}_\psi$. We therefore coarse-grain out any other information about~$\ket{\psi}$ (or the theory) by defining an equivalence relation~$\sim$, where~$\ket{\psi_1} \sim \ket{\psi_2}$ if~$O_{\psi_1}(X_i)$ and~$O_{\psi_2}(X_i)$ have the same singularity structures.  Then by construction, if~$\ket{\psi_1} \sim \ket{\psi_2}$, both~$\ket{\psi_1}$ and~$\ket{\psi_2}$ give rise to the same higher dimensional conformal geometry.  The \textit{causal state} of~$\ket{\psi}$ is defined as the equivalence class of~$\sim$ to which~$\ket{\psi}$ belongs, which we denote as~$\widetilde{\ket{\psi}}$.  Thus by construction, the causal states that give rise to well-defined dual conformal geometries are in one-to-one correspondence with those conformal geometries.

We now proceed to formalize this construction.  Consider the Hilbert space~$\Hcal$ of the QFT and the space~$\overline{\Lcal(\Hcal)}$ of all linear operators on~$\Hcal$\footnote{This space consists of  \textit{all} linear operators, and not just bounded ones; hence the notation~$\overline{\Lcal(\Hcal)}$.}.  A state~$\ket{\psi} \in \Hcal$ induces a map to the extended complex numbers~$\overline{\cnum}$ (also known as the Riemann sphere), corresponding to the complex numbers plus the ``point at infinity'':
\be
\ev{\cdot}_\psi: \overline{\Lcal(\Hcal)} \to \overline{\cnum},
\ee
where~$\ev{A}_\psi \equiv \me{\psi}{A}{\psi}$  for any operator~$A \in\overline{\Lcal(\Hcal)}$. This map is just the usual expectation value (though note that in general~$A$ need not be Hermitian).

Consider next any local operator~$\Ocal(X) \in \overline{\Lcal(\Hcal)}$; we use its composition to define a binary map on~$(d+3)$ copies of the QFT spacetime manifold~$\partial M$ (denoted as~$\partial M^{d+3}$) as follows\footnote{For CFTs, $\partial M$ should be taken to be in a standard conformal frame~\cite{EngHor15} to avoid issues with boundary geodesic incompleteness.}:
\begin{subequations}
\label{eq:Wmap}
\bea
W_{\psi}: \partial M^{d+3} &\to \{0,1\}\\
(X_1,\ldots,X_{d+3}) &\mapsto \begin{cases} 1 & \mbox{if } \left\|\left ( \prod\limits_{i=1}^{d+3}\partial_{i}^{k_{i}} \right) \ev{\Ocal(X_1) \cdots \Ocal(X_{d+3})}_\psi\right\| = \infty \mathrm{\ for \ some \ } \{k_{i}\} \\ 0 & \mbox{otherwise} \end{cases}
\eea
\end{subequations}
where the norm is taken with respect to the metric on~$\partial M$ (in a standard conformal frame),~$\infty$ denotes the ``point at infinity'' of~$\overline{\cnum}$ (i.e.~the north pole of the Riemann sphere), and the $k_{i}$ are integers. The simplest example of such singularities is when all the $k_{i}$ vanish: i.e. when the correlator diverges. More generally, these singularities correspond to nonanalyticities in the correlator, which are captured by derivatives. The support of this map (that is, those points~$X_{i}$ on which~$W_{\psi} = 1$) consists of state-independent singularities (short-distance and boundary light-cone singularities) and state-dependent singularities (e.g.~bulk-sourced singularities).   It is these latter singularities that define the lightcone cuts, and therefore the emergent causal geometry, when one exists.  The map~$W$ thus precisely captures all of the information necessary to reconstruct the causal geometry, should it exist.

It is therefore natural to use this map to define the equivalence relation~$\sim$ by requiring that the maps~$W_{\psi_1}$ and~$W_{\psi_2}$ of any two equivalent states~$\ket{\psi_1}$ and~$\ket{\psi_2}$ coincide:
\begin{defn}
Let~$\ket{\psi_1}$ and~$\ket{\psi_2}$ be two states in the same Hilbert space with associated maps~$W_{\psi_1}$,~$W_{\psi_2}$.  Then we say that~$\ket{\psi_1}$ and~$\ket{\psi_2}$ are \textit{causally equivalent}, written as~$\ket{\psi_1} \sim \ket{\psi_2}$, if we have
\be
W_{\psi_1}(X_1,\ldots,X_{d+3}) = W_{\psi_2}(X_1,\ldots,X_{d+3})
\ee
for all~$\{X_i\}_{i=1}^{d+3}$.
\end{defn}

In principle, the relation~$\sim$ may depend on the choice of operator~$\Ocal$ used to define the map~$W_{\psi}$.  However, the states in which we are ultimately interested are those with the property that the corresponding maps~$W_{\psi}$ are independent of choice of~$\Ocal$ (so that the same bulk conformal geometry arises for all of them); that is, whenever $\Ocal$ has a corresponding dual field in the bulk, which interacts perturbatively.  We therefore do not bother labeling~$\sim$ by a choice of~$\Ocal$, though we will comment more on this issue in Section~\ref{sec:disc}.

Note that the construction reviewed in Section~\ref{subsec:cutreview} made use of the lightcone cuts, rather than the map~$W_{\psi}$, to reconstruct the conformal geometry.  Why do we not define the equivalence relation~$\sim$ using the cuts directly?  The reason is that the discussion in Section~\ref{subsec:cutreview} uses statements originating from the bulk geometry, while here we are interested in defining the equivalence~$\sim$ in a purely field theoretic manner, applicable to all states whether or not they have a semiclassical holographic dual.  Indeed, given an \textit{arbitrary} state~$\ket{\psi}$, the cut construction from the map~$W_{\psi}$, as  outlined in Section~\ref{subsec:cutreview}, need not yield any well-defined surfaces.  In such a case, the state~$\ket{\psi}$ does not have a semiclassical holographic dual.  Thus though its map~$W_{\psi}$ exists independently of the holographic dual, it is meaningless to talk about its ``lightcone cuts''.  However, in the case where~$\ket{\psi}$ \textit{does} have a well-defined semiclassical dual, the construction of Section~\ref{subsec:cutreview} is guaranteed to give rise to well-defined ``cuts'' of~$\ket{\psi}$.  In such a case, the map~$W_{\psi}$ gives rise to a unique set of cuts, and any two such states~$\ket{\psi_1}$,~$\ket{\psi_2}$ with the same~$W_{\psi_{1,2}}$ must necessarily give rise to the same cuts, and therefore the same dual conformal geometry.  The upshot is that working with~$W_{\psi}$ rather than the (potentially ill-defined) ``cuts'' makes the equivalence relation~$\sim$ a well-defined purely field-theoretic object.

This precise definition of the equivalence relation~$\sim$ therefore gives us a precise field-theoretic definition of the causal state~$\widetilde{\ket{\psi}}$ as the equivalence class of~$\ket{\psi}$.  It is natural to ask whether this equivalence relation is nontrivial; that is, whether there exist different states that are equivalent under this relation.  If this were not the case, then  states~$\ket{\psi}$ could be uniquely specified by the singularity structure of the correlator~$\ev{\Ocal(X_1) \cdots \Ocal(X_{d+3})}_\psi$. However, states are specified uniquely by the collection of expectation values of \textit{all} operators, not by the singularity structure of just one. This is a general expectation, but there is a clear argument that the equivalence relation cannot be trivial\footnote{We thank Gary Horowitz for this argument.}: in the large-$N$ limit of AdS/CFT, we may perturb a state at $\mathcal{O}(N^{0})$; this results in subleading corrections that do not change the singularity structure, but by construction change the state. Therefore multiple states must belong in the same equivalence causal equivalence class.

This picture is of course perfectly consistent with the dual geometric interpretation: from the point of view of a holographic bulk dual, a particular state~$\ket{\psi}$ is dual to a bulk geometry, while a causal state~$\widetilde{\ket{\psi}}$ is dual to the causal structure of the bulk geometry.  The coarse-graining from~$\ket{\psi}$ to~$\widetilde{\ket{\psi}}$ thus simply corresponds to a coarse-graining over all possible bulk conformal factors.

\subsection{Reduced Causal States}
\label{subsec:reduced}

An important aspect of holography is subregion/subregion duality~\cite{BouLei12}: while it is valuable to work with the  dual of the entire field theory, we now understand that  \textit{subregions} of the field theory have sensible bulk subregion duals~\cite{JafLew15, DonHar16}.  This property of holography is no less interesting in the present context, as it has applications to an open question in AdS/CFT: what is the dual of the causal wedge of some boundary region?

We now develop some technology that will allow us to provide an answer in Section~\ref{sec:causalwedge}. To that end, note that the equivalence relation~$\sim$ was defined by the condition that $\ket{\psi_1} \sim \ket{\psi_2}$ if and only if the maps~$W_{\psi_1}$ and~$W_{\psi_2}$ agree on the entire CFT spacetime.  There are two natural generalizations of this equivalence: one in which we require only some of the~$X_i$ in~$W_\psi(X_i)$ to lie in~$\Rcal$, and another stronger version in which all the~$X_i$ are contained to lie in~$\Rcal$.  We will discuss the first generalization here; the second generalization has interesting ties to quantum error correction, and we postpone a more thorough investigation of it to Section~\ref{sec:QEC}.

Roughly, we define a ``reduced'' equivalence relation~$\sim_\Rcal$ by the condition that~$\ket{\psi_1} \sim_\Rcal \ket{\psi_2}$ if and only if the corresponding cuts of~$\ket{\psi_1}$ and~$\ket{\psi_2}$ agree within~$\Rcal$ regardless of what they may do elsewhere, thereby ``coarse-graining'' over any behavior of the cuts outside of the region~$\Rcal$ as shown in Figure~\ref{fig:rcs}. To make this precise, we define the equivalence relation~$\sim_\Rcal$ as follows:

\begin{figure}[t]
\centering
\includegraphics[width=3.5cm,page=6]{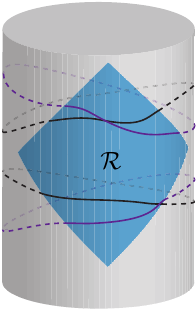}
\caption{For the reduced causal density matrix of~$\Rcal$, only the structure of the lightcone cuts inside~$\Rcal$ matters: lightcone cuts that agree on~$\Rcal$ and do not agree elsewhere are identical as perceived by the reduced causal density matrix on~$\Rcal$. The dashed lines denote those portions of the cuts over which we coarse-grain.}
\label{fig:rcs}
\end{figure} 

\begin{defn}
\label{defn:causalequivR}
Let~$\Rcal$ be some subregion of the boundary spacetime, and let~$\ket{\psi_1}$ and~$\ket{\psi_2}$ be two states with associated maps~$W_{\psi_1}$,~$W_{\psi_2}$.  Then we say that~$\ket{\psi_1}$ and~$\ket{\psi_2}$ are \textit{causally equivalent on~$\Rcal$}, written as~$\ket{\psi_1} \sim_\Rcal \ket{\psi_2}$, when for every~$k = 2,\ldots,d+3$ 
\be
W_{\psi_1}(X_1,\ldots,X_k,Y_{k+1},\ldots,Y_{d+3}) = W_{\psi_2}(X_1,\ldots,X_k,Y'_{k+1},\ldots,Y'_{d+3})
\ee
for all~$\{X_i\}_{i=1}^k \subset \Rcal$ such that at least two of the~$X_i$ are chronally related and for some~$\{Y_i\}_{i=k+1}^{d+3}$,~$\{Y'_i\}_{i=k+1}^{d+3} \not\subset \Rcal$.  
\end{defn}

It follows easily from this definition that the equivalence relations~$\sim_\Rcal$ admit a nesting structure: if~$\ket{\psi_1} \sim_\Rcal \ket{\psi_2}$, then~$\ket{\psi_1} \sim_{\Rcal'} \ket{\psi_2}$ for any~$\Rcal' \subset \Rcal$.  In particular, if~$\Rcal$ is the entire boundary spacetime~$\partial M$, then~$\sim_{\partial M}$ is just the equivalence relation~$\sim$ defined in the previous section.  Thus if~$\ket{\psi_1} \sim \ket{\psi_2}$, then~$\ket{\psi_1} \sim_\Rcal \ket{\psi_2}$ for any region~$\Rcal$.

Now, let us define \textit{causal states on~$\Rcal$}, denoted as~$\widetilde{\ket{\psi}}_\Rcal$, as the equivalence classes of~$\sim_\Rcal$.  The nesting structure of the equivalence relations~$\sim_\Rcal$ implies the following property: given a causal state~$\widetilde{\ket{\psi}}_\Rcal$ and a subregion~$\Rcal' \subset \Rcal$, we can obtain the causal state~$\widetilde{\ket{\psi}}_{\Rcal'}$ by taking the equivalence class of~$\widetilde{\ket{\psi}}_\Rcal$ under~$\sim_{\Rcal'}$; however, given a causal state~$\widetilde{\ket{\psi}}_{\Rcal'}$, it is not sensible to take an equivalence class under~$\sim_\Rcal$ to reconstruct~$\widetilde{\ket{\psi}}_\Rcal$.  Thus, by going from~$\widetilde{\ket{\psi}}_\Rcal$ to~$\widetilde{\ket{\psi}}_{\Rcal'}$, information about the state is lost; this is the precise interpretation of the coarse-graining advertised above.  If~$\Rcal' \subset \Rcal$, then~$\widetilde{\ket{\psi}}_{\Rcal'}$ is more coarse-grained than~$\widetilde{\ket{\psi}}_\Rcal$.

\subsection{A Causal Hilbert Space and Causal Density Matrix}
\label{subsec:causalspace}

So far, the equivalence relations~$\sim_\Rcal$ serve a useful formal role to identify precisely what boundary information is sufficient to reconstruct a bulk causal geometry corresponding to the causal wedge of~$\Rcal$.  However, it is natural to wonder whether there is some additional structure associated with this equivalence relation; for instance, does it induce a quotient space structure of the Hilbert space~$\Hcal$?

To investigate this question, consider~$\Hcal$ as a vector space (that is, let us temporarily ignore the inner product structure on~$\Hcal$).  Recall that an equivalence relation induces a quotient of~$\Hcal$ if and only if it is a \textit{congruence} on~$\Hcal$; that is, if it is compatible with the vector space structure of~$\Hcal$.  In other words,~$\sim_\Rcal$ will induce a quotient space of~$\Hcal$ if and only if for any~$\ket{\psi}$,~$\ket{\psi'}$,~$\ket{\phi}$,~$\ket{\phi'} \in \Hcal$, 
\be
\label{eq:congruence}
\mbox{If } \ket{\psi} \sim_\Rcal \ket{\psi'} \mbox{ and } \ket{\phi} \sim_\Rcal \ket{\phi'} \mbox{ then } \ket{\psi+\phi} \sim_\Rcal \ket{\psi'+\phi'}.
\ee
We argue that the above relation does indeed hold, although we do not offer a proof.  First, note that we have
\begin{subequations}
\bea
O_{\psi + \phi}(X_{i}) &= O_{\psi}(X_{i}) + O_{\phi}(X_{i}) + \me{\psi}{\Ocal(X_1)\cdots \Ocal(X_{d+3})}{\phi} + \me{\phi}{\Ocal(X_1)\cdots \Ocal(X_{d+3})}{\psi}, \\
O_{\psi' + \phi'}(X_{i}) &= O_{\psi'}(X_{i}) + O_{\phi'}(X_{i}) + \me{\psi'}{\Ocal(X_1)\cdots \Ocal(X_{d+3})}{\phi'} + \me{\phi'}{\Ocal(X_1)\cdots \Ocal(X_{d+3})}{\psi'}.
\eea
\end{subequations}
Clearly the singularity structures of the first two terms in each line above must agree within~$\Rcal$ (in the sense of Definition~\ref{defn:causalequivR}), since~$\ket{\psi} \sim_\Rcal \ket{\psi'}$ and~$\ket{\phi} \sim_\Rcal \ket{\phi'}$.  Thus~$\sim$ will be a congruence if the mixed matrix elements~$\me{\phi}{\Ocal(X_1)\cdots \Ocal(X_{d+3})}{\psi}$ do not contain any singularities besides those present in~$O_\psi$ and~$O_\phi$.

To gain some intuition for these cross terms from AdS/CFT, consider the case where~$\ket{\psi}$ and~$\ket{\phi}$ have semiclassical duals\footnote{We thank Gary Horowitz for providing us with the following argument.}, and denote the energies of these states by~$E_\psi$ and~$E_\phi$.  Because we are necessarily working in the large-$N$ limit, these energies will generically be of order~$N^2$; thus~$\ket{\psi}$ will be composed of a superposition of energy eigenstates within a small energy window~$(E_\psi - \Delta, E_\psi + \Delta)$, where~$\Delta$ is subleading in~$1/N$, and likewise for~$\ket{\phi}$.  If~$E_\psi \neq E_\phi$, these energy windows will not overlap, and thus~$\ket{\phi}$ and~$\ket{\psi}$ must be orthogonal:~$\inprod{\psi}{\phi} = 0$ (to leading order in~$1/N$).  But since~$Ocal(X_1)\cdots \Ocal(X_{d+3})$ consists of order unity copies of the local operator~$\Ocal(X)$, the energy of the state~$Ocal(X_1)\cdots \Ocal(X_{d+3})\ket{\phi}$ must also be~$E_\phi$ to order~$N^2$, and thus by the same logic, the inner product~$\me{\phi}{\Ocal(X_1)\cdots \Ocal(X_{d+3})}{\psi} = 0$ as well.  Thus in this case, the matrix elements do not add any additional singularities to the correlator, as desired.

In the case that~$\ket{\psi}$ and~$\ket{\phi}$ have the same energy or do not have semiclassical duals, or more generally if we work outside the context of AdS/CFT, we have less intuition for the behavior of the cross terms~$\me{\phi}{\Ocal(X_1)\cdots \Ocal(X_{d+3})}{\psi}$.  Nevertheless, it seems reasonable to expect that the singularity structure of these matrix elements should arise only from the singularity structures of~$O_\psi$ and~$O_\phi$.  If this expectation is borne out, these cross terms cannot add any new singularities, and therefore the singularity structures of~$O_{\psi + \phi}$ and~$O_{\psi' + \phi'}$ must match within~$\Rcal$\footnote{The roughness of this argument means we are necessarily ignoring the possibility of fine-tuned cases where the singularities cancel.}.  This again verifies that~$\sim_\Rcal$ is a congruence of~$\Hcal$.

If the rough arguments outlined above are correct, then the equivalence relation~$\sim_\Rcal$ defines a quotient vector space~$\widetilde{\Hcal}_{\Rcal} \equiv \Hcal/\!\sim_\Rcal$, which we naturally dub the \textit{causal vector space of~$\Rcal$} (or just the \textit{causal vector space} when~$\Rcal$ is the entire QFT spacetime).  Thus the causal states~$\widetilde{\ket{\psi}}_\Rcal$ are not just equivalence classes, but can be thought of as elements of the vector space~$\widetilde{\Hcal}_\Rcal$.  Due to the nesting structure of the~$\sim_\Rcal$, this family of vector spaces obeys the property that for any~$\Rcal' \subset \Rcal$,~$\widetilde{\Hcal}_\Rcal$ can be quotiented by~$\sim_{\Rcal'}$ to obtain~$\widetilde{\Hcal}_{\Rcal'}$ (but not the other way around).  Again, this is an instance of coarse-graining under exclusion.

Moreover, any operator~$A \in \overline{\Lcal(\Hcal)}$ that obeys the property
\be
A\ket{\psi} \sim_\Rcal A\ket{\psi'} \mbox{ for all } \ket{\psi}, \ket{\psi'} \mbox{ such that } \ket{\psi} \sim_\Rcal \ket{\psi'}
\ee
gives rise to an operator~$\widetilde{A}_\Rcal$ on~$\widetilde{\Hcal}_\Rcal$.  Finally, note that we can augment the vector space~$\widetilde{\Hcal}_\Rcal$ with an inner product to obtain a causal inner product space\footnote{This is due to the fact that an inner product can be introduced on any vector space (over~$\mathbb{R}$ or~$\cnum$) via the introduction of a Hamel basis.  However, an inner product on~$\widetilde{\Hcal}_\Rcal$ is not inherited in any natural way from the inner product on~$\Hcal$, since given any~$\ket{\psi}$,~$\ket{\psi'}$,~$\ket{\phi}$,~$\ket{\phi'}$ in~$\Hcal$ such that~$\ket{\psi} \sim_\Rcal \ket{\psi'}$ and~$\ket{\phi} \sim_\Rcal \ket{\phi'}$, it need not be the case that~$\inprod{\psi}{\phi} = \inprod{\psi'}{\phi'}$.}.  We expect that the resulting inner product space will be complete (since~$\Hcal$ was) and is therefore a Hilbert space; however, subtleties arising from infinite sums may spoil the completeness property. If this is the case, $\widetilde{\Hcal}_\Rcal$ is nevertheless a normed vector space.  Assuming the above considerations are correct (and that the resulting space is complete), we conclude that~$\widetilde{\Hcal}_\Rcal$ is a \textit{causal Hilbert space}, and we will term it thus throughout.

This inner product structure thereby allows us to define covectors, and thereby define a \textit{causal density matrix on~$\Rcal$} as
\be
\label{eq:CDM}
\tilde{\rho}_\Rcal \equiv \widetilde{\ket{\psi}}_\Rcal \widetilde{\bra{\psi}}_\Rcal.
\ee
This causal density matrix is equivalent to the original state~$\widetilde{\ket{\psi}}_\Rcal$.

As an aside, we should note that there is another sensible notion of what could be meant by a ``causal density matrix''.  So far, we have restricted ourselves to pure states~$\ket{\psi}$ because they are elements of the Hilbert space~$\Hcal$, and therefore the causal state~$\widetilde{\ket{\psi}}_\Rcal$ is an element of the causal Hilbert space~$\widetilde{\Hcal}_\Rcal$.  However, we may also be interested in mixed states, which are defined instead by a density matrix~$\rho$, which is a linear operator on~$\Hcal$.  The map~$W$ defined in~\eqref{eq:Wmap} can then be defined using~$\rho$ by replacing the expectation value with~$\left\langle A \right\rangle_\rho \equiv \Tr(\rho A)$ for any~$A \in \overline{\Lcal(\Hcal)}$.  This again defines an equivalence relation~$\sim_\Rcal$ by Definition~\ref{defn:causalequivR}, which may then be used to define an alternative causal density~$\tilde{\rho}'_\Rcal$ matrix as the equivalence class of~$\sim_\Rcal$ to which~$\rho$ belongs.  However, this definition is less appealing because unlike~$\widetilde{\ket{\psi}}_\Rcal$, it is not clear that the object~$\tilde{\rho}'_\Rcal$ can be thought of as an element of some structured ``causal'' vector space like~$\widetilde{\Hcal}_\Rcal$, and unlike~$\tilde{\rho}_\Rcal$, it is not an operator on the quotient space~$\widetilde{\Hcal}_\Rcal$.  For these reasons, we do not make use of~$\tilde{\rho}'_\Rcal$ here.


\section{Causal Quantum Error Correction}
\label{sec:QEC}

\begin{figure}[t]
\centering
\includegraphics[width=3.5cm,page=7]{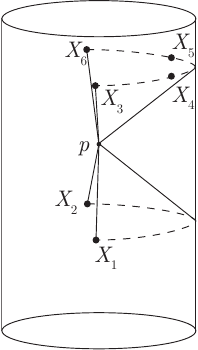}
\caption{In order to reconstruct the conformal metric at some bulk point~$p$ from singularities of the~$(d+3)$-point correlator~$O_{\psi}(X_i)$, the points~$X_i$ need to be sufficiently spread out around the boundary to ensure momentum conservation at~$p$.}
\label{fig:bndryangle}
\end{figure}

The definition of the equivalence relation~$\sim_\Rcal$ given above consisted of a ``partial'' restriction of the map~$W_{\psi}$ to the region~$\Rcal$ in the sense that only some of the points~$x_i$ were required to be in~$\Rcal$.  From the perspective of the lightcone cuts, this condition is necessary to conserve momentum at all bulk points whose cuts intersect~$\Rcal$: the set of boundary points that render~$O_{\psi}(X_i)$ singular must span some minimum ``angle'', as shown in Figure~\ref{fig:bndryangle}. Defining the reduced causal state~$\widetilde{\ket{\psi}}_\Rcal$ in this way ensures that it captures the lightcone cuts that intersect~$\Rcal$, as we may use points outside of~$\Rcal$ to ensure energy-momentum conservation for obtaining them.

We may, however, consider a stricter approach wherein we rely on nothing but data within $\Rcal$ itself, as is more parallel to notions of subregion/subregion duality in AdS/CFT.  Let us therefore define a different causal state which captures this intuition. Consider restricting the map~$W_{\psi}$ \textit{entirely} to~$\Rcal$; that is, constrain \textit{all} the~$X_i$ to be contained in~$\Rcal$.  Roughly, this constraint allows us to recover the open subsets of only those lightcone cuts that intersect $\Rcal$ on sufficiently large sets, i.e.~on sets that are sufficiently large to ensure energy-momentum conservation in the bulk\footnote{This restriction applies at least via the correlator construction; other ways of obtaining the lightcone cuts, should they exist, may circumvent this issue.}, and thus restricts the subset of the bulk that can be reconstructed.  Increasing the size of~$\Rcal$ increases the maximum ``angle'' that can be spanned by points contained within~$\Rcal$, and thus allows the recovery of the conformal metric of progressively larger subsets of the bulk.

\begin{figure}[t]
\centering
\subfigure[]{\includegraphics[width=4.5cm,page=8]{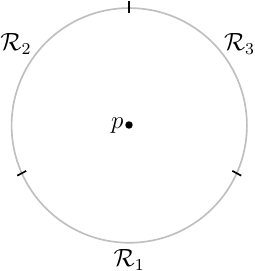}
\label{subfig:birdview}
}
\hspace{2cm}
\subfigure[]{\includegraphics[width=4cm,page=9]{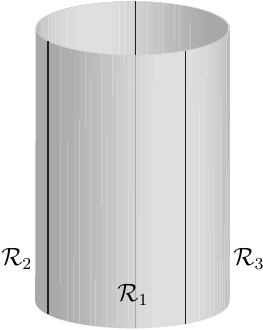}
\label{subfig:timelikestrips}
}
\caption{The boundary is divided into three timelike strips: $\Rcal_{1}, \Rcal_{2}, \Rcal_{3}$. If all the~$X_i$ are restricted to any one of the boundary subregion~$\Rcal_{i}$, the conformal metric cannot be reconstructed at all bulk points in causal contact with~$\Rcal_{i}$, as the corresponding Landau diagram does not conserve energy momentum.  However, increasing the size of~$\Rcal_{i}$ to include both $\Rcal_{1}$ and $\Rcal_{2}$ allows the reconstruction of the conformal metric at points ``deeper'' in the bulk.} 
\label{fig:increaseR}
\end{figure}

This feature -- wherein bulk data that is not contained in a boundary region~$\Rcal'$ can instead be recovered from a sufficiently larger region~$\Rcal \supset \Rcal'$ -- is strongly reminiscent of quantum error correction and quantum secret sharing~\cite{AlmDon15}.  We can make this connection clearer by formulating the observations above more precisely, starting with the following definition:

\begin{defn}
\label{defn:strongcausalequivR}
Let~$\Rcal$ be some subregion of the boundary spacetime, and let~$\ket{\psi_1}$ and~$\ket{\psi_2}$ be two states with associated maps~$W_{\psi_1}$,~$W_{\psi_2}$.  Then we say that~$\ket{\psi_1}$ and~$\ket{\psi_2}$ are \textit{strongly causally equivalent on~$\Rcal$}, written as~$\ket{\psi_1} \hat{\sim}_\Rcal \ket{\psi_2}$, when
\be
W_{\psi_1}(X_1,\ldots,X_{d+3}) = W_{\psi_2}(X_1,\ldots,X_{d+3})
\ee
for all~$\{X_i\}_{i=1}^k \subset \Rcal$.
\end{defn}

This equivalence is just a much stronger version of that in Definition~\ref{defn:causalequivR}; we may use it  analogously to define the \textit{strongly reduced causal state on~$\Rcal$} as the equivalence class~$\widehat{\widetilde{\ket{\psi}}}_\Rcal$ of~$\ket{\psi}$ under~$\hat{\sim}_\Rcal$.  Importantly, if~$\ket{\psi}$ has a semiclassical bulk dual, then~$\widehat{\widetilde{\ket{\psi}}}_\Rcal$ only contains sufficient information to reconstruct the conformal metric at those bulk points~$p$ that are null-separated from~$(d+3)$ points in~$\Rcal$ by an allowed Landau diagram.  More precisely, we define what is meant by a sufficient region~$\Rcal$ as follows:

\begin{defn}
Consider a state~$\ket{\psi}$ with a well-defined semiclassical holographic dual causal structure, and consider a point~$p$ in this dual geometry.  An open connected boundary region~$\Rcal$ is \textit{sufficient} for reconstructing the conformal metric at~$p$ if it contains at least~$(d+3)$ points null-separated from~$p$ by an allowed Landau diagram.  Otherwise,~$\Rcal$ is \textit{insufficient} for reconstructing the conformal metric at~$p$.
\end{defn}

Note that in principle a set of~$(d+3)$ points null-separated from~$p$ (by an allowed bulk Landau diagram) is only sufficient to define the point~$p$, not to reconstruct the conformal metric at~$p$.  However, because~$\Rcal$ is required to be open, the existence of a set of~$(d+3)$ points null-separated from~$p$ in fact implies that~$\Rcal$ contains enough points to reconstruct open intervals of the lightcone cut of~$p$ and the points in the neighborhood of~$p$.  These in turn are sufficient to reconstruct the conformal metric at~$p$, justifying the above definition.

The relationship to quantum error correction arises when we consider constructing sufficient regions from a union of insufficient ones.  Let $p$ be a bulk point as before and consider a region~$\Rcal$ sufficient to reconstruct the conformal metric at~$p$.  We may divide~$\Rcal$ into a number of disjoint regions~$\Rcal_i$, each of which is insufficient for reconstructing the conformal metric at~$p$; for simplicity, let us consider a case where~$\Rcal$ is the full boundary spacetime, which is divided into three strips, each of which is insufficient, but such that the union of any two is sufficient, as shown in Figure~\ref{fig:increaseR}.  Then by the above definitions, access to the strongly reduced causal state on the union~$\Rcal_{ij} \equiv \Rcal_i \cup \Rcal_j$ of any two regions is sufficient to fix the conformal metric at~$p$.

We thus find that bulk reconstruction via bulk point singularities has the following two properties:
\begin{enumerate}
	\item Quantum Error Correction: the conformal metric at~$p$ is protected against erasure of any one of the~$\Rcal_i$;
	\item Quantum Secret Sharing: any one of the~$\Rcal_i$ does not contain knowledge of the conformal metric at~$p$. 
\end{enumerate}
Here by ``erasure of~$\Rcal_i$'', we mean restriction of the map~$W_{\psi}$ to the complement of~$\Rcal_i$ in the sense of Definition~\ref{defn:strongcausalequivR}.

The comparison with the usual notion of quantum error correction is striking.  Recall that in the typical density-matrix based quantum error correction, the Hilbert space can be factorized via three causal diamonds~$\Rcal_i$ as~$\Hcal = \Hcal_1 \otimes \Hcal_2 \otimes \Hcal_3$, and a state~$\rho$ gives rise to reduced density matrices~$\rho_i$ on each of these three diamonds.  ``Erasure'' in this case refers to the loss of one of the three diamonds (say~$\Rcal_3$), and quantum error correction is manifested in the fact that a bulk observable can be reconstructed from the reduced density matrix of the remaining regions~($\rho_{12}$).  Quantum secret sharing refers to the fact that the reduced density matrix of a single diamond is insufficient to reconstruct this bulk observable.

The discussion in almost unchanged in the present case, except that instead of reduced density matrices, we use the strongly reduced causal states.  Then quantum error correction refers to the fact that the strongly reduced causal state of the union of two of the three strips is sufficient to reconstruct the conformal metric at~$p$, and quantum secret sharing is manifested in the fact that the strongly reduced causal state of only one strip is not.

\section{The Dual to the Causal Wedge}
\label{sec:causalwedge}

\begin{figure}[t]
\centering
\includegraphics[width=3.3cm,page=10]{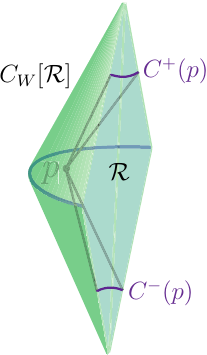}
\caption{The portion of the lightcone cuts that intersect a region~$\Rcal$ correspond to those bulk points $p$ in~$J^+[\Rcal] \cap J^-[\Rcal]$; when~$\Rcal$ is a causal diamond, this region is precisely the causal wedge~$C_{W}[\Rcal]$.}
\label{fig:causalwedgepoints}
\end{figure}

Let us now address the bulk interpretation of the (reduced) causal state~$\widetilde{\ket{\psi}}_\Rcal$.  By construction,~$\widetilde{\ket{\psi}}_\Rcal$ contains the information both necessary and sufficient to reconstruct those portions of the lightcone cuts of~$\ket{\psi}$ (when they exist) that intersect~$\Rcal$.  These cuts correspond precisely to those bulk points in~$I^+[\Rcal] \cap I^-[\Rcal]$, as shown in Figure~\ref{fig:causalwedgepoints}.  In particular, when~$\Rcal$ is a causal diamond, this region is just the causal wedge~$C_W[\Rcal]$ of~$\Rcal$.  Since open subsets of the cuts corresponding to points in~$C_W[\Rcal]$ can be used to reconstruct the conformal metric in~$C_W[\Rcal]$, we find that the reduced causal state~$\widetilde{\ket{\psi}}_\Rcal$ contains precisely the field theoretic information necessary and sufficient to reconstruct the conformal geometry of~$C_W[\Rcal]$!

It is worth reminding the reader that part of our purpose in this paper is the proposition for a shift in the general approach to the causal wedge; in particular, because the causal wedge is defined in terms of the causal structure of the bulk geometry, we have no right to expect that anything beyond its causal structure should have a ``nice'' field theoretic dual.  For the causal structure itself, this dual is precisely~$\widetilde{\ket{\psi}}_\Rcal$: \textit{the reduced causal state is the field-theoretic dual of (the causal structure of) the causal wedge}.

Our position is motivated in part by the general arguments of~\cite{EngWal17}, which suggest that the area of the causal surface has no simple information-theoretic interpretation.  This is in contrast with the story regarding the HRT surface and the entanglement wedge; there the reduced density matrix~$\rho_\Rcal$ associated to some causal diamond~$\Rcal$ is understood to be the field-theoretic dual to the entanglement wedge, and the area of the HRT surface is dual to the von Neumann entropy of~$\rho_\Rcal$.

Nevertheless, it is possible to draw some parallels between the reduced density matrix~$\rho_\Rcal$ and our causal state~$\widetilde{\ket{\psi}}_\Rcal$.  The idea is to focus on the notion of coarse-graining: the reduced density matrix~$\rho_\Rcal$ is obtained from a state~$\rho$ by tracing out the degrees of freedom in the complement~$\overline{\Rcal}$ of~$\Rcal$:
\be
\rho_\Rcal = \Tr_{\overline{\Rcal}} \rho.
\ee
The von Neumann entropy of~$\rho_\Rcal$, also called the entanglement entropy of~$\Rcal$, is a measure of the coarse-graining performed in going from~$\rho$ to~$\rho_\Rcal$, and corresponds to the information lost in restricting the state to the region~$\Rcal$.

Our causal state~$\widetilde{\ket{\psi}}$ is similarly obtained from~$\ket{\psi}$ via a coarse-graining over any information not contained in the map~$W_{\psi}$.  Is there number associated to this coarse-graining that quantifies the information lost in going from~$\ket{\psi}$ to~$\widetilde{\ket{\psi}}_\Rcal$?  Likewise, is there a number than quantifies the information lost when going from~$\widetilde{\ket{\psi}}_\Rcal$ to~$\widetilde{\ket{\psi}}_{\Rcal'}$ for $\Rcal'\subset \Rcal$?  We suggest answers to these questions below.

\subsubsection*{A Measure of Coarse-Graining}

Roughly speaking, the quantification of the coarse-graining performed in going from a state~$\ket{\psi}$ to a causal state~$\widetilde{\ket{\psi}}_\Rcal$ requires a measure of the number of unique physical states in the Hilbert space~$\Hcal$ that map to the same causal state~$\widetilde{\ket{\psi}}_\Rcal$.  A na\"ive such measure would be the cardinality of the causal state~$\widetilde{\ket{\psi}}_\Rcal$, thought of as a coset of the equivalence relation~$\sim_\Rcal$.  However, a more natural measure can be obtained by exploiting the structure of the equivalence relation~$\sim_\Rcal$ on the Hilbert space~$\Hcal$.

To that end, note that the property~\eqref{eq:congruence} from Section~\ref{subsec:causalspace} implies that for a given $\ket{\psi}$, the set of vectors
\be
\{\ket{\phi} \in \Hcal : \ket{\phi} \sim_\Rcal \ket{\psi} \}
\ee
is closed under addition and scalar multiplication.  By augmenting the above set with the zero vector~$\ket{0}$\footnote{We emphasize that this is the zero vector, \textit{not} the vacuum state.}, we obtain a vector space:
\be
\Hcal_{\Rcal,\psi} \equiv \{\ket{0}\} \cup \{\ket{\phi} \in \Hcal : \ket{\phi} \sim_\Rcal \ket{\psi} \}.
\ee
The dimension~$D_{\Rcal,\psi} \equiv \mathrm{dim}(\Hcal_{\Rcal,\psi})$ is a well-defined object, and captures precisely the notion of how much coarse-graining has been performed in obtaining the causal state.  However, note that it is constructed from the state in a radically different way than the entanglement entropy is; this difference is consistent with our claim that the causal and entanglement wedges are fundamentally different.

Of course, because~$\Hcal$ is infinite-dimensional, it is possible for~$D_{\Rcal,\psi}$ to be infinite-dimensional as well.  In such a case, we imagine rendering~$\Hcal$ finite via the usual introduction of some kind of regulator.

\subsubsection*{Conformal Invariants in the Causal Wedge}

Which \textit{bulk} objects are natural candidates for obtaining $\mathbb{C}$-numbers from the (reduced) causal state? Bulk considerations automatically lead us to consider different objects that interestingly turn out to be finite and independent of renormalization scheme.  Consider a state~$\ket{\psi}$ known (by the procedure outlined in Section~\ref{subsec:cutreview}) to have a semiclassical holographic dual, and consider a boundary causal diamond~$\Rcal$.  We may consider using the causal state~$\widetilde{\ket{\psi}}_\Rcal$ to define numbers that encode properties of~$\widetilde{\ket{\psi}}_\Rcal$, regardless of the coarse-graining performed in obtaining this state from~$\ket{\psi}$.  Conceivably, such objects should have bulk duals constructed from the conformal geometry of the causal wedge of~$\Rcal$.

We may therefore consider bulk objects of the form
\be
\label{eq:Qdef}
Q = \int_B \bm{f},
\ee
where~$B$ is some bulk surface or region associated to~$\Rcal$ and~$\bm{f}$ is a form (of appropriate dimension) that is left unchanged by Weyl transformations of the bulk geometry.  Note that the causal holographic information, which takes the same form as the integral above with~$B$ the causal surface and~$\bm{f}$ taken to be the natural volume form on~$B$, is \textit{not}  Weyl invariant, since the natural volume form on~$B$ is not.

In general it is unclear whether objects of the form~\eqref{eq:Qdef} have a clear field theoretic interpretation.  However, in the special case where~$B$ is taken to be the causal wedge of~$\Rcal$, the integral above has an interpretation as an integral over the space of all cuts that intersect~$\Rcal$ (since each point in~$C_{W}[\Rcal]$ corresponds to a pair of cuts intersecting~$\Rcal$).  Even in this special case, however, it is not clear if there is a natural choice of~$\bm{f}$: in even bulk dimensions, it is known that there are many allowed choices of local~$\bm{f}$, but no local Weyl-invariant~$\bm{f}$ exists in odd bulk dimensions~\cite{FefGra85,BaiEas94,FefGra07,GraHir08} (indeed, this is the reason there is no conformal anomaly in odd-dimensional CFTs).  It may be the case that the objects~\eqref{eq:Qdef} are just uninteresting, or that~$\bm{f}$ needs to be constructed in some unusual way in odd dimensions.

\subsubsection*{Causal RG Flow and Bulk Depth}

We have so far been using the term ``coarse-graining'' in a broad sense to refer to the process of decreasing our knowledge about a state.  However, it is worth asking if there is any relationship between causal states and coarse-graining in the strict sense of renormalization group (RG) flow: that is, can the causal density matrix be related to a flow from the UV to the IR?

\begin{figure}[t]
\centering
\includegraphics[width=7cm,page=11]{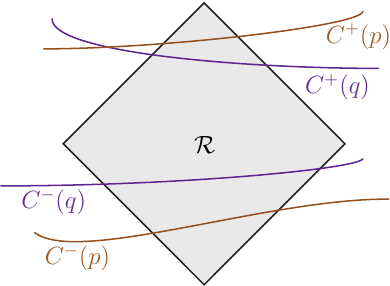}
\caption{Adapted from~\cite{Eng16}. The lightcone cuts of two points~$p$ and~$q$ give a well-defined notion of relative bulk depth:~$p$ is deeper into the bulk that~$q$ relative to some region~$\Rcal$ if and only if the light-cone cuts of~$p$ in~$\Rcal$ sandwich those of~$q$.}
\label{fig:bulkdepth}
\end{figure}

It has been recently shown by one of us in~\cite{Eng16} that bulk depth can be defined covariantly via the lightcone cuts: a point~$p$  was defined as deeper than a point~$q$ relative to a boundary subregion~$\Rcal$ if the lightcone cuts of~$p$ ``sandwich'' the lightcone cuts of~$q$, as shown in Figure~\ref{fig:bulkdepth}. Motion into the bulk was shown to correspond to both larger distances and longer time scales on the boundary, providing a precise way of relating RG flow to bulk depth. Since this approach towards bulk depth uses the lightcone cuts, our construction of the causal density matrix seems naturally related to it.  Is RG flow defined by the lightcone cuts an RG flow of the causal density matrix?

The answer is yes. It was shown in~\cite{Eng16} that defining bulk depth using lightcone cuts is equivalent to defining bulk depth using the causal wedge:~$p$ is deeper than~$q$ if every causal wedge containing~$p$ also contains~$q$.  This could have already been seen as a foreshadowing that the structure of lightcone cuts would give the dual to the causal wedge. An equivalent statement, using our causal state formulation, is that $p$ is deeper than $q$ if and only if every reduced causal state $\widetilde{\ket{\psi}}_\Rcal$ that can be used to reconstruct the conformal metric at~$p$ contains sufficient information to reconstruct the conformal metric at~$q$.  Thus RG flow in the sense of a flow ``deeper into the bulk'' corresponds directly to a flow among reduced causal states over progressively larger boundary causal diamonds.



\section{General Holography}
\label{sec:hol}

Since quantum gravity is expected to be holographic in broad generality, it is natural to ask which, if any, of our results are also valid beyond strong holography.  In addressing this question, it is useful to use some kind of explicit general holographic framework as a guide.  A convenient option is Bousso's covariant formulation of general holography~\cite{Bou99b,Bou99c,Bou99d,Bou02}.  Let us therefore examine how our results generalize to Bousso's holographic framework, bearing in mind that the lessons we learn may apply even to dualities not described by it.

The essence of Bousso's formalism~\cite{Bou99c} is the identification of distinguished hypersurfaces in a spacetime, termed \textit{preferred holographic screens}, that are particularly well-suited for general holography.  These preferred holographic screens are defined as codimension-one hypersurfaces that admit a foliation into spacelike codimension-two surfaces~$\sigma(r)$ (where~$r$ indexes the foliation) called leaves, such that at least one of the two null expansions at each~$\sigma(r)$ vanishes:
\be
\theta(\sigma(r))=0.
\ee
By the Bousso bound~\cite{Bou99b}, the area (or more generally, the generalized entropy~\cite{BouFis15}) of distinguished cross-sections of these hypersurfaces bounds the entropy of their interior.  This feature makes preferred holographic screens distinguished hypersurfaces for general holography.  It is especially appealing that the AdS boundary is itself a very special type of preferred holographic screen (a so-called optimal screen).  Generically, however, preferred holographic screens are not necessarily timelike, and they need not be restricted to the asymptotic boundary. 

The class of preferred holographic screens has since been broken down into two categories: past and future holographic screens~\cite{BouEng15a, BouEng15b}, which we shall refer to as ``screens'' for short.  A past screen is characterized by the property that the two future-directed null expansions~$\theta_k(\sigma(r))$ and~$\theta_l(\sigma(r))$ of its leaves~$\sigma(r)$ are zero and strictly positive, respectively:
\begin{subequations}
\bea
& \theta_k(\sigma(r))=0, \\
&\theta_l(\sigma(r))>0.
\eea
\end{subequations}
A future screen is defined analogously, with~$\theta_l(\sigma(r))$ strictly negative.

Now, consider a screen in a semiclassical spacetime that arises from some underlying theory of quantum gravity.  It has been suggested that the interior of this screen, which must itself be described by perturbative quantum gravity, is holographically dual to some theory that can be embedded on the screen~\cite{NomSal16, NomSal16b}.  What type of theory is this?

Because screens have distinguished spatial slices, it is immediately clear that a theory living on a screen and dual to its interior will not generally be relativistic.  Moreover, the leaves of screens obey an area law~\cite{BouEng15a, BouEng15b} (or more generally, the Generalized Second Law~\cite{BouEng15c}), implying that the entropy of the dual theory living on the screen is not constant.  This in turn would appear to require violations of unitarity; see~\cite{Moo17} for a recent discussion of these issues.  When the screen in question is null, there are subtleties: the leaves may have constant area, and the leaves are not distinguished slices.  It is possible to define a theory on the screen using light-front quantization, but this approach is nonrelativistic.  Finally, while there are examples of holographic screens of timelike signature, a generic holographic screen has mixed signature, alternating (in a way constrained by~\cite{BouEng15b}) between spacelike, timelike, and null. 

These considerations all indicate that general holography as formulated on holographic screens does not obey our definition of strong holography\footnote{The \textit{only} exceptions are conformal boundaries and null surfaces with vanishing expansion; any spatial cross section of these is a marginally trapped surface, allowing for the possibility that strong holography is obeyed.}.  At first this might seem an insurmountable hurdle to generalizing any of the work in this paper to holographic screens. However, the situation is not so dire: recall that the conditions of weak holography require that (\ref{H1}) the \textit{bulk} quantum fields are well-approximated by local, relativistic quantum field theory;~(\ref{H2prime}) the boundary theory is well-defined on a timelike or null geometry; and~(\ref{H3prime}) \textit{some} version of the extrapolate dictionary holds (correlators may map to more exotic objects, for example).  It then follows that the lightcone singularities of the bulk correlator must be dual to a singularity in some observable in the dual theory.  That is, if~$\ev{\phi(x_{1})\cdots \phi(x_{d+3})}$ limits to~$O(X_{1},\ldots, X_{d+3})$ when the $x_{i}\rightarrow X_{i}$, then our formalism still applies as long as~$O$ represents a quantity that is computable in the boundary theory.

In particular, if such an~$O(X_1,\ldots,X_{d+3})$ exists, then the lightcone cuts would be obtained by demanding that~$O(X_{1},\ldots, X_{d+3})$ be singular as two of the points $X_{i}$ are varied, thus tracing out a cut.  By the propositions of Section~\ref{subsec:cutreview}, this cut will be a complete spacelike slice of any timelike or null connected past/future holographic screen.  The procedure for obtaining algebraic equations for the conformal metric remains unaltered. Thus lightcone cuts of timelike or null holographic screens exist and have the requisite properties for the emergence of conformal geometry. 

Similarly, almost all of the other results of Section~\ref{sec:CDM} generalize.  Since the boundary theory in question is either non-local, non-relativistic, non-unitary, or all of the above, it is not clear what the appropriate notion of a state is from the boundary perspective.  However, under the assumption that there exists \textit{some} way of defining a boundary state, we may again identify any two states of the system under the equivalence relation~$\sim$ constructed from the object~$O$, thus allowing us to define causal states, reduced causal states, and causal density matrices\footnote{It is not clear in this case whether structure of the original space of states of the boundary theory will be inherited by the space of causal states, so the existence of an object like the causal Hilbert space~$\widetilde{{\cal H}}$ is not guaranteed in general.}.  Interestingly, the results of Section~\ref{sec:QEC} apply as well, suggesting that quantum gravity beyond AdS/CFT is generally quantum error correcting.

We thus find that a well-defined theory living on a timelike or null space can have a semiclassical conformal geometry dual with an extrapolate dictionary as defined by Property~\ref{H3prime} if and only if there exists a quantity~$O(X_{1},\ldots, X_{d+3})$ such that the singularities of~$O$ give rise to a family of lightcone cuts, which in turn result in a system of algebraic equations with a nontrivial solution.  In fact, when this is the case, this solution \textit{defines} the extrapolate dictionary.

However, there remains a significant hurdle to applying this formalism to holographic screens in broad generality.  In Section~\ref{subsec:cutreview}, we presented a generalization of the theorems in~\cite{EngHor16a} to include timelike and null boundaries that are not necessarily asymptotic boundaries (and when they are, they need not be asymptotically AdS).  However, holographic screens are often spacelike, and generally have mixed signature.  For instance, in the context of dS/CFT, the asymptotic timelike infinity of dS is an optimal screen that is spacelike.  The proofs of Propositions~1-4 do not obviously generalize to such cases, but based on the fact that lightcone cuts may be defined in any spacetime, we expect that generalizations of these propositions should exist.  If so, the procedure of obtaining the conformal geometry from singularities of~$O$ could be extended to arbitrary holographic screens.  This would be an interesting and important question to address.

\section{Discussion}
\label{sec:disc}

In this paper, we have defined causal states and causal density matrices, new constructs in local quantum field theory.  As long as the strong holography conditions~\ref{H1}-\ref{H3} are obeyed, the causal density matrix simultaneously provides~\textit{(i)} a sufficient and necessary condition for the existence of a semiclassical holographic dual causal structure to a state, and~\textit{(ii)} an explicit procedure for obtaining the conformal geometry of this dual.  The construction of the causal state involved coarse-graining a regular state over everything except the minimum information required to define a causal, (approximately) local holographic dual (when one exists).  Under the strong holography conditions~\ref{H1}-\ref{H3}, we have argued that the space of causal states is a Hilbert space.

Our definition allows us to define not just the causal states associated to the full boundary theory, but to also restrict to a subregion~$\Rcal$.  We performed this restriction in two ways: a full restriction to~$\Rcal$ exhibits features typical of quantum error correction and quantum secret sharing, while a partial restriction to~$\Rcal$ yields the dual of the causal wedge.  We emphasize that these statements are not limited to AdS/CFT, but rather constitute general results about the states of QFTs with semiclassical holographic duals.  This, coupled with the holographic hypothesis, has the far-reaching implication that quantum error correction is a feature of semiclassical states of quantum gravity with a holographic extrapolate dictionary. 

Within the AdS/CFT correspondence, we have found that the reduced causal density matrix provides a natural answer to a longstanding question: what is the dual of the causal wedge?  As recent arguments in~\cite{EngWal17} have shown that superficial parallels between the entanglement and causal wedges are misleading rather than instructive, we have proposed a new, alternative approach to the dual of the causal wedge.  The crux of this approach is the observation that the causal wedge itself is defined purely in terms of the bulk conformal geometry, and therefore the primary innate property of the causal wedge is precisely its causal structure.  The only natural field-theoretic dual is thus one that is sensitive \textit{only} to the bulk conformal geometry; this is precisely the reduced causal density matrix.  Our argument thus implies that the natural dual to the causal wedge should be ignorant of the conformal factor, the bulk matter fields, and the bulk dynamics, all of which play no part in the definition of the causal wedge.

Since (under the conditions of strong holography) we have argued that the space of causal states is a Hilbert space, we have exploited this structure to define a measure~$D_{\Rcal,\psi}$ of the amount of coarse-graining performed in going from a physical state~$\ket{\psi}$ to a reduced causal state~$\widetilde{\ket{\psi}}_\Rcal$; this is essentially a measure of the size of the space of states causally equivalent to~$\ket{\psi}$.  We have also suggested that any objects constructed from the causal state alone should have a bulk interpretation as conformal invariants, though a detailed analysis of what these objects might be is beyond the scope of this paper.

While our discussion of causal states (and causal density matrices) and their bulk duals has focused on states of QFTs, we have argued that many of our results continue to hold in the context of weak holography, where conditions~\ref{H1},~\ref{H2prime}, and~\ref{H3prime} hold.  Importantly, this generalization relaxes the requirement that the boundary theory be a QFT and allows it to live on a null as well as timelike geometry.  A full generalization to holographic theories on arbitrary spaces, and in particular to ones with spacelike boundaries, would be a natural next step, and would also be a significant step towards developing an understanding of the emergence of time.

Since our primary purpose here was to define causal states and point out their importance in the reconstruction of semiclassical bulk duals, our analysis of their applications was necessarily brief and exploratory.  Many interesting questions remain; we list a few below.

\paragraph{Hilbert Space Factorization:} While we have established that parallels between the causal and entanglement pictures are misleading, it is nonetheless interesting to ask if the causal Hilbert space structure permits the definition of analogues of objects akin to entanglement or R\'enyi entropies from the causal density matrix.  A natural way to attempt to define such objects is via the factorization of the causal Hilbert space in some way to obtain a definition of a partial trace of the causal density matrix.  While it is unlikely that the causal Hilbert space factorizes over subregions like the physical Hilbert space does, it is possible that there exists a different, natural factorization that permits the definition of a partial trace over some part of the causal Hilbert space.

\paragraph{Multiple Dual Geometries:} We defined the causal state using~$(d+3)$-point correlators of a local operator~$\Ocal(X)$.  By construction, the emergent causal structure sourced by the singularities of $\Ocal(X)$ is such that high energy quanta of the field $\phi$ dual to $\Ocal$ (i.e. the characteristics of $\phi$) are null. It is interesting to note the possibility that two different local operators~$\Ocal_1(x)$ and~$\Ocal_2(x)$ give rise to two \textit{independent} sets of cuts, each of which gives rise to a different well-behaved conformal geometry.   In this case, it would appear that \textit{both} conformal geometries are emergent from the same boundary theory state. There are two ways in which this might happen: first, if the high energy quanta of two bulk fields $\phi_{1}$ and $\phi_{2}$ have different propagation velocities\footnote{We thank David Simmons-Duffin for pointing this out.}, then the two apparently distinct conformal geometries capture different manifestations of dynamics on the same background.  In such cases, there should exist a consistent way of embedding both fields $\phi_{1}$ and $\phi_{2}$ on the same geometry: the boundary state is really dual to one geometry. The second possibility is that there is no way of embedding the fields $\phi_{1}$ and $\phi_{2}$ in the same conformal geometry. In this latter case, we would say that the boundary state genuinely sources multiple bulk geometries. Can this happen?

\begin{figure}[t]
\centering
\includegraphics[width=3.3cm,page=12]{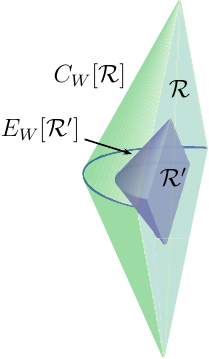}
\caption{The entanglement wedge $E_{W}[\Rcal']$ is a proper subset of the causal wedge $C_{W}[\Rcal]$. The reduced causal density matrix of $\Rcal$ completely fixed the conformal geometry of $E_{W}[\Rcal']$. Thus, the reduced causal density matrix of $\Rcal$, which is defined using time-separated points, must significantly constrain the reduced density matrix of $\Rcal'$, which is defined spatially. }
\label{fig:wedgenesting}
\end{figure}

\paragraph{Relationship to Entanglement:} Consider some boundary causal diamond~$\Rcal$ and a subset of it~$\Rcal' \subset \Rcal$ such that the entanglement wedge of~$\Rcal'$ lies within the causal wedge of~$\Rcal$: $E_{W}[\Rcal']\subset C_{W}[\Rcal]$, as shown in Figure~\ref{fig:wedgenesting}. Then the interior geometry of~$E_{W}[\Rcal']$ is fixed up to an overall function (i.e.~the conformal factor) by the reduced causal density matrix of~$\Rcal$.  It is thus clear that the causal density matrix significantly constrains the reduced density matrix.  This is somewhat surprising, however: the causal density matrix is by definition dependent on temporally-separated points, while the regular density matrix is not. This is quite likely related to the fact that the entanglement wedge is defined locally while the causal wedge is defined teleologically.  Thus while we have argued that the causal wedge and entanglement wedge should be thought of in fundamentally different ways, it may be that the causal density matrix will allow purely field theoretic comparisons with the reduced density matrix. This could also potentially answer an RG flow question: we could give a definition of bulk depth via entanglement wedge inclusion instead of causal wedge inclusion (the latter of which is equivalent to the definition of bulk depth via the lightcones cuts~\cite{Eng16}). We would expect that one definition is constrained by the other, however it is not clear what the precise relation is in the bulk. It is possible that this is more easily done by comparing the causal and regular density matrices.


\paragraph{Pathological Geometries:} Some of the results presented in Section~\ref{subsec:cutreview} required the use of global or AdS hyperbolicity.  This requirement, however, was used only for convenience; it is not required for the reconstruction of the conformal metric from open subsets of the lightcone cuts.  This observation raises an interesting question: if the lightcone cut construction can be used to reconstruct non-globally hyperbolic or otherwise pathological spacetimes (e.g.~ones containing closed timelike curves), can these pathologies be interpreted from a boundary perspective?  In other words, is there some structure to the singularities of~$O_\psi(X_i)$ which is indicative of a pathological dual causal geometry?  Presumably, the corresponding boundary states should be pathological from a purely field theoretic perspective; thus answering this question would provide a boundary interpretation of causally pathological bulk duals.

\section*{Acknowledgements}

It is a pleasure to thank Chris Akers, Raphael Bousso, Chris Fewster, Daniel Harlow, Tom Hartman, Gary Horowitz, Bernard Kay, Juan Maldacena, Don Marolf, Mark Mezei, Eric Perlmutter, David Simmons-Duffin, Arkady Tseytlin, Mark Van Raamsdonk, and Aron Wall for useful discussions and correspondence.  The work of NE was supported in part by NSF grant PHY-1620059.  SF was supported by STFC grant ST/L00044X/1, and wishes to thank the UCSB physics department for hospitality while some of this work was completed.


\bibliographystyle{jhep}
\bibliography{all}
\end{spacing}
\end{document}